\documentclass[12pt,preprint]{aastex}
\def\degpoint{\ifmmode ^{\rm{o}}\!. \else $^{\rm{o}}\!.$\fi}

\newcommand{\ms}{\mbox{m\,s$^{-1}$}}

\newcommand{\Msun}{\mbox{M$_{\odot}$}}

\newcommand{\Mjup}{\mbox{M$_{\rm Jup}$}}

\newcommand{\Mearth}{\mbox{M$_{\oplus}$}}

\usepackage{subfigure}
\begin{document}

\title{Pursuing the planet-debris disk connection: Analysis of upper 
limits from the Anglo-Australian Planet Search}

\author{Robert A.~Wittenmyer\altaffilmark{1,2}, Jonathan 
P.~Marshall\altaffilmark{1,2} }

\altaffiltext{1}{School of Physics, UNSW Australia, Sydney 2052, 
Australia}
\altaffiltext{2}{Australian Centre for Astrobiology, UNSW Australia, 
Sydney 2052, Australia}
\email{
rob@phys.unsw.edu.au}

\shorttitle{Planet-debris disk connection}
\shortauthors{Wittenmyer \& Marshall}

%-------------------------------------------------------------------
\begin{abstract}

\noindent Solid material in protoplanetary discs will suffer one of two 
fates after the epoch of planet formation; either being bound up into 
planetary bodies, or remaining in smaller planetesimals to be ground 
into dust. These end states are identified through detection of 
sub-stellar companions by periodic radial velocity (or transit) 
variations of the star, and excess emission at mid- and far-infrared 
wavelengths, respectively. Since the material that goes into producing 
the observable outcomes of planet formation is the same, we might expect 
these components to be related both to each other and their host star. 
Heretofore, our knowledge of planetary systems around other stars has 
been strongly limited by instrumental sensitivity. In this work, we 
combine observations at far-infrared wavelengths by \textit{IRAS}, 
\textit{Spitzer}, and \textit{Herschel} with limits on planetary 
companions derived from non-detections in the 16-year Anglo-Australian 
Planet Search to clarify the architectures of these (potential) 
planetary systems and search for evidence of correlations between their 
constituent parts.  We find no convincing evidence of such 
correlations, possibly owing to the dynamical history of the disk 
systems, or the greater distance of the planet-search targets.  Our 
results place robust limits on the presence of Jupiter analogs which, in 
concert with the debris disk observations, provides insights on the 
small-body dynamics of these nearby systems.

\end{abstract}


%--------------------------------------------------------------------
\section{Introduction}

Circumstellar debris discs around main-sequence stars are composed of 
second-generation dust produced by the attrition of larger bodies 
\citep{BacPar1993}, which are remnants of primordial protoplanetary 
discs \citep{Hernandez2007}.  Exoplanets form early in the history of 
these systems, with $\sim$10\,\Mearth\ planets required to capture a gas 
envelope from the protoplanetary disc before it dissipates 
\citep[typically within 3-10~Myr][]{Hernandez2007,Ribas2014}, whereas 
terrestrial planet formation by hierarchical growth \citep{lissauer95, 
pollack96} may continue over longer timescales of a few tens of Myr 
\citep[e.g. Earth-Moon forming collision at 
$\sim$~40~Myr][]{Canup2008,Canup2012}.

Planet formation requires the hierarchical growth of dust grains to 
pebbles and thereafter to larger bodies eventually ending up at 
asteroids and comets - the planetesimals from which exoplanets form 
\citep{Perryman2011, Armitage2013}.  At the same time, collisions 
between these planetesimals produce the dust grains we observe as the 
visible components of debris discs. Since planetesimals are key to 
production of both planets and dusty debris, one might expect the 
properties of planets and debris around a star to be mutually dependent.  
This expectation has been strengthened by the direct imaging of several 
exoplanet systems around debris disc host stars and indirectly by the 
structural features observed in many debris discs (warps, off-sets, 
asymmetries), which are often inferred to be due to the gravitational 
perturbation of the debris by one or more unseen exoplanet(s) \citep[see 
reviews by][]{Wyatt08,Krivov10,Moromartin13}.

Observations of nearby sun-like stars at far-infrared wavelengths by the 
\textit{Spitzer} \citep{Werner2004} and \textit{Herschel} 
\citep{Pilbratt2010} space telescopes revealed evidence for several 
correlations between planets and debris.  \textit{Herschel} observations 
determined that a higher statistical incidence of exoplanets around 
debris disk host stars is seen, explicitly linking these two components 
of planetary systems \citep{Bryden2013}.  Further correlations between 
the presence of both a debris disc and low-mass planets around low 
(sub-solar) metallicity stars \citep{Wyatt2012,Marshall2014}, and 
between the presence of a debris disc and cold (distant) Jovian 
planet(s) \citep{Maldonado2012}, have also been identified.  Such 
correlations can be understood, and even expected, within a picture of 
planet formation via core accretion, and the subsequent dynamical 
interaction between planet(s) and planetesimal belts.

Previous studies to determine the underlying connections between these 
distinct components of planetary systems -- stars, planets and debris -- 
have concentrated on analysis of target samples consisting of known 
exoplanet host stars with or without debris 
\citep{Beichman2006,Moromartin2007,Kospal2009,Bryden2009}.  Little 
regard was given to the potential that any given star may in fact host a 
planet below the threshold of current detection capabilities.  In this 
work we take account of the threshold upper limits for companions 
orbiting stars targeted by the Anglo-Australian Planet Search, some of 
which are already known to host planetary companions.

The Anglo-Australian Planet Search (AAPS) is an ongoing, 16-year 
high-precision radial-velocity survey of $\sim$250 nearby solar-type 
stars \citep{tinney01, butler01}.  It has achieved a consistent velocity 
precision of 2-3\ms\ for its lifetime, making the AAPS a world leader in 
the detection of long-period planets analogous to Jupiter.  These 
``Jupiter analogs'' are among the more recent of the $\sim$40 planets 
discovered by the AAPS, \citep[e.g.][]{142paper, 2jupiters}, and are the 
focus of ongoing observations and simulation work \citep{jupiters, 
witt13}.  In addition to the planet discoveries, the AAPS data are 
useful for setting limits on the presence of undetected planets 
\citep[e.g.][]{otoole09, foreverpaper, etaearth}.

In this paper, we use the AAPS data, including non-detections, to 
further explore the connection between debris disks and planets.  For 
the first time, we include the detection limits for targets which have 
debris disc observations but as yet no known planets.  In Section 2, we 
describe the sample and detection-limit technique.  Section 3 presents 
the results, and we give our conclusions in Section 4.

%--------------------------------------------------------------------
\section{Observations and the stellar sample }

The stellar physical parameters used in the analysis, i.e. luminosity, 
photospheric temperature, age, and metallicity, were taken from 
\cite{Takeda2007} and \cite{Valenti2005}. The distances were taken from 
the re-reduction of the \textit{HIPPARCOS} catalogue by 
\cite{Vanleeuwen2007}.

\subsection{Far-infrared observations}

Of the 141 AAPS stars analysed here, 54 were observed at far-infrared 
wavelengths by the \textit{Herschel} Space Observatory 
\citep{Pilbratt2010} with the Photodetector Array Camera and 
Spectrometer instrument \citep[PACS;][]{Poglitsch2010,Balog2013} through 
a combination of the Guaranteed Time debris discs programme, the Open 
Time Key Programmes ``Disc Emission via a Bias-free Reconnaissance in 
the Infrared /Submillimetre'' \citep[DEBRIS;][]{Matthews2010} and ``DUst 
around NEarby Stars'' \citep[DUNES;][]{Eiroa2013}, and the Open Time 
programmes ``Search for Kuiper-belts Around Radial-velocity Planet 
Stars'' \citep[SKARPS;][]{Bryden2013,Kennedy2013} and program 
OT1\_amoromar\_1 (PI: A. Moro-Mart\'in).  A further 11 stars were 
observed at 70~$\mu$m by the \textit{Spitzer} Space Telescope 
\citep{Werner2004} with its ``Multi-band Imaging Photometer for 
Spitzer'' instrument \citep[MIPS;][]{Gordon2007}.  Finally, upper limits 
at 60\,$\mu$m were taken from the \textit{IRAS} Faint Source Catalogue 
for an additional 39 targets.  A total of 104/141 AAPS stars therefore 
have some measure of the presence (or absence) of debris in their 
circumstellar environment.  Of these, 21 stars have detected 
infrared excesses.

\textit{Herschel} flux densities were taken from the literature where 
available \citep{Lestrade2012,Eiroa2013,Marshall2014}.  For targets 
without published measurements, the PACS data were reduced and analysed 
using the \textit{Herschel} Interactive Processing Environment 
\citep[HIPE;][]{ott10} using the standard data reduction scripts and 
following the method laid out in \citet{Eiroa2013}.  \textit{Spitzer} 
flux densities were taken from the literature, namely 
\citet{Trilling2008} and \citet{Bryden2009}.

\subsection{Dust limits}

Dust fractional luminosities, or upper limits in the case of 
non-detection at far-infrared wavelengths, of the AAPS target stars were 
calculated from fitting of a modified blackbody \citep{Wyatt08} to the 
\textit{Spitzer} MIPS measurements at 70~$\mu$m \citep[compiled 
from][]{Trilling2008,Bryden2009} and \textit{Herschel} PACS measurements 
at 70 and/or 100 and 160~$\mu$m, along with optical, near- and 
mid-infrared and sub-mm measurements (where available) taken from the 
literature, following the approach of \cite{Marshall2014}.  In the case 
of targets with only upper limits on their emission at far-infrared 
wavelengths, a dust temperature of 50 K was assumed for the fitting 
process.  We compare the AAPS results presented here with results from 
the \textit{Herschel}-observed radial-velocity planet host sample from 
\cite{Marshall2014} (see Figure 2).

%For targets without sub-mm measurements, a wavelength of 
%$\lambda_{0}\!=\!210~\mu$m and $\beta\!=\!1$ were assumed for the break 
%in the blackbody fit to the disk's emission.

%We compare the AAPS results presented here with results from the 
%literature for known debris disc host stars with sub-millimeter flux 
%densities taken from \citep{NajWil2005,WilAnd2006,Nilsson2010,Panic2013} 
%and the \textit{Herschel}-observed radial-velocity planet host sample 
%from \cite{Marshall2014} (see Figure 2).

\subsection{Planetary detection limits}

Amongst the 141 AAPS stars examined here, 43 are previously known to 
host one or more companions.  For the targets known to host planets, we 
fit for and removed the signals of those planets, making use of 
additional velocity data from the literature where available.  The stars 
considered here and the characteristics of the velocity data are given 
in Table~\ref{rvdata}.  The detection limit was determined by adding a 
fictitious Keplerian signal to the data, then attempting to recover it 
via a generalized Lomb-Scargle periodogram \citep{zk09}.  Here, we have 
assumed circular orbits; for each combination of period $P$ and 
radial-velocity semiamplitude $K$, we tried 30 values of orbital phase.  
The radial-velocity amplitude $K$ of the injected planet is increased 
until 99\% of orbital configurations at a given $P$ are recovered with a 
false-alarm probability \citep{ss10} of less than 1\%.  This approach is 
virtually identical to that used in our previous work 
\citep[e.g.][]{limitspaper, wittenmyer09, jupiters}.  Because this 
approach uses a periodogram to determine detectability, gaps in data 
sampling can conspire to make some trial periods fail to produce a 
significant peak even for very large amplitudes ($K_{max}=200$\ms).  
These situations result in allegedly undetectable trial signals 
\citep{foreverpaper}.  We compensate for this artefact in the following 
way: If an injected signal at a trial period $P$ fails to be detected 
for all $K$ values, the algorithm switches from the periodogram to the 
$F$-test, starting again at $K_{min}=1$\ms\ and increasing $K$ until all 
trial configurations at that $(P,K)$ combination result in an rms which 
differs from the original data at the 99\% significance level.  The 
upper limits on planetary companions were computed in this way for all 
141 stars.  We choose the detection limit at 5\,AU as a representative 
figure of merit, being the approximate orbital radius of Jupiter, and 
being at an orbital period ($\sim$12 yr) within the total timespan of 
the AAPS data (16 yr)\footnote{For the median stellar mass 1\Msun, 
$a_{5AU}\sim$12 yr. The range in this sample is 0.45-1.72\Msun, or 
$a_{5AU}\sim$ 16.9-8.5 yr, respectively)}.

%--------------------------------------------------------------------
\section{Results and Discussion}

\begin{figure}
\plotone{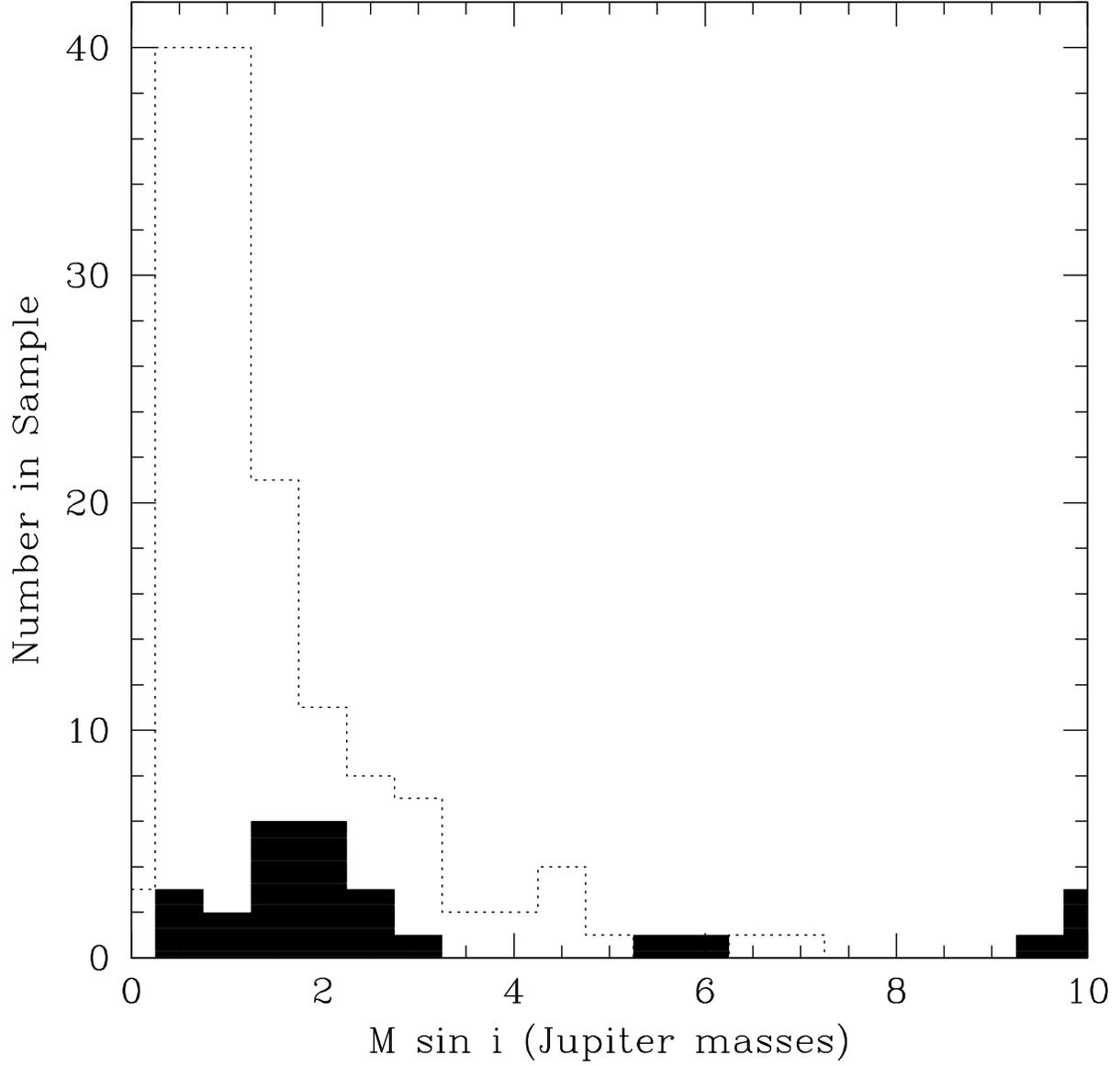}
\caption{Dashed histogram: planetary detection limits at 5\,au for the 
141 stars in this sample. Filled histogram: known planets in the sample 
with $a>1$\,au and m~sin~$i>$0.2\Mjup. }
\label{planetlimits}
\end{figure}

\begin{figure}
\includegraphics[width=0.6\textwidth,angle=90]{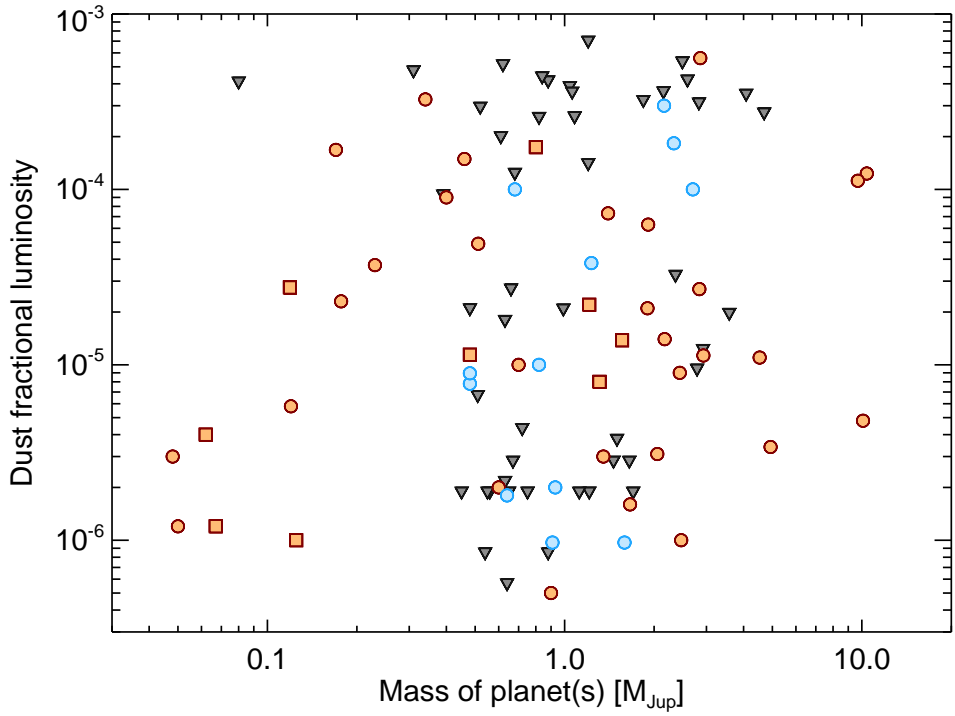}
\caption{Dust fractional luminosity as a function of the total mass of 
planets in the system.  Triangles: upper limits for both dust and 
planets.  Blue circles: stars with dust disks and no known planets.  
Red circles: stars with both detected dust and planets.  Red squares: 
stars with planets but no dust. }
\label{fig1}
\end{figure}

Planetary detection limits in the sample of 141 stars are shown as a 
histogram in Figure~\ref{planetlimits}; for comparison, known planets 
with $a>1$\,au and m~sin~$i>$0.2\Mjup\ are shown as a filled histogram.  
In Figure \ref{fig1}, the AAPS stars are presented as coloured or black 
data points, while those taken from literature sources are in greyscale.  
Broadly speaking, two distinct regions dominate the parameter space 
illustrated in Fig.~2: systems with high fractional luminosities and no 
known planets (of any given age) occupy the top left (with masses 
derived from sub-mm flux densities), whilst systems with low dust 
luminosities (or only upper limits) and massive Jovian exoplanets occupy 
the bottom right.  This is suggestive that massive, cool Jovian planets 
preclude a peaceful coexistence with a debris belt 
\citep{Maldonado2012}. Exceptions to this trend do exist including 
HD\,95086 and HR\,8799 with bright discs and massive exoplanets at 
comparable orbital radii 
\citep[][e.g.]{rameau13a,rameau13b,moor13,marois08,marois10,matthews14}, 
but such cases are usually young ($<$100 Myr) and A-type stars.  For a 
summary of A-star debris population statistics, see e.g. \citet{su06} 
and \citet{thureau14}.  Such systems are quite different from the 
targets of the AAPS survey which are mature ($t_{\rm age}~>~1$Gyr), 
sun-like (FGK type) stars.  For a summary of debris around FGK stars, 
see e.g. \citet{Bryden2009}, \citet{Maldonado2012}, and 
\citet{Eiroa2013}. 

In the search for any correlation between the planets and debris 
it should be noted that only the brightest ends of the distributions of 
both planet mass (modulo orbital radius/instrument sensitivity) and dust 
luminosity have heretofore been measured.  That we are able to discern 
any correlation between these components of planetary systems, however 
tentatively, is perhaps unexpected 
\citep{Maldonado2012,Wyatt2012,Marshall2014}.  It is as yet impossible 
to detect direct dusty analogues to our solar system, due to the 
faintness of its cool debris disc, the Edgeworth-Kuiper belt, 
\citep[$L_{\rm IR}/L_{\star}~\sim~1.2\times10^{-7}$,][]{vitense12}.  
Long term radial-velocity programs monitoring exoplanet host stars are 
still only in their third decade, with the AAPS being the longest 
currently running exoplanet survey, although the now defunct Lick survey 
still holds the record for longest duration \citep{fischer14}.  The 
solar system was speculated to be at the fainter end of the dust 
brightness distribution \citep[amongst the bottom 10\%][]{greaves10}, 
although subsequent \textit{Herschel} observations suggest it may in 
fact lie close to the average disk brightness (Moro-Martin et al., in 
prep.).

\subsection{Statistical Analysis} 

Recent analysis of results from \textit{Spitzer} \citep{Maldonado2012, 
Wyatt2012} and \textit{Herschel} \citep{Bryden2013,Marshall2014} have 
identified correlations between the presence of debris and exoplanets 
around sun-like stars, finding that debris discs are more common around 
stars with known planetary companions \citep{Bryden2013} and that 
low-mass planet hosts favour the presence of debris over those stars 
with Jovian-mass companions \citep{Wyatt2012,Marshall2014}.  In this 
section, we apply the Kolmogorov-Smirnov (KS) and Fisher exact tests to 
the sample of AAPS stars presented here, looking for similar 
correlations.

We consider the dust fractional luminosities, or 3-$\sigma$ upper 
limits, derived from either \textit{Spitzer} MIPS 70~$\mu$m or 
\textit{Herschel} PACS 100~$\mu$m measurements assuming a disc 
temperature of 50~K, with a preference for \textit{Herschel} values if 
both are available due to the superior angular resolution of the PACS 
instrument over that of MIPS.  The mass upper limits at 5~au for the 
stars are derived in this work.  For the purposes of statistical 
analysis in this work, any star with a mass limit $<~1~M_{\rm Jup}$ and 
no known Jovian-mass companion is ruled to be a potential low-mass 
planet host star, whilst those stars with mass limits $\geq~1~M_{\rm 
Jup}$ (or host a known Jovian-mass companion) may potentially harbour a 
Jupiter analogue planet and are therefore potential high-mass planet 
hosts in this analysis.

In the AAPS sample presented here there is a total of 141 stars.  
Amongst these, 43 stars host known substellar 
companions\footnote{HD\,164427 is a brown dwarf, \citep{tinney01}.} of 
which nine also host a cool debris disk.  Of the remainder, twelve stars 
are known to host a debris disk without exhibiting any evidence of 
planets, whilst the remaining 87 stars in the sample have no 
observational evidence of a companion planetary system.

\begin{figure*}
\plotone{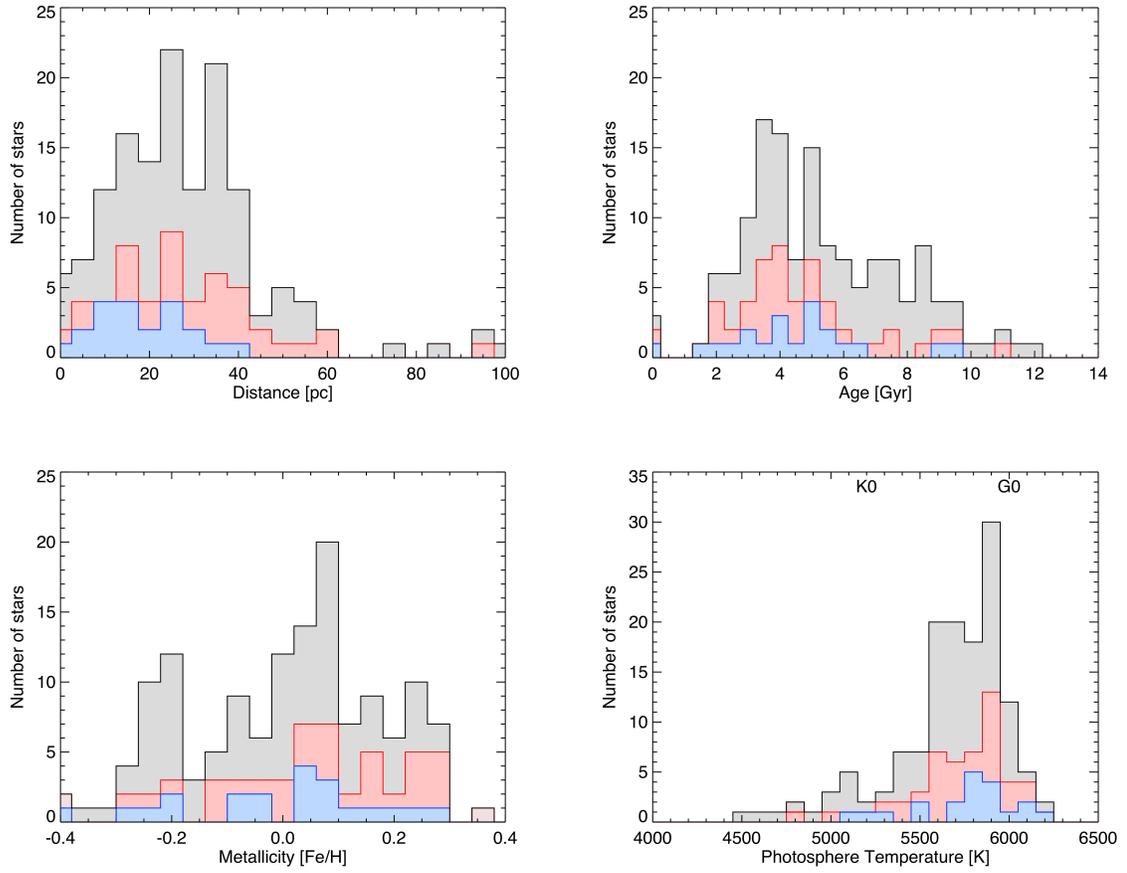}
\caption{Distribution of the distance, age, metallicity, and $T_{eff}$ 
for the 141 AAPS stars considered here.  Grey histogram: total sample.  
Blue histogram: stars with detected debris ($N=21$).  Red histogram: 
stars with known planets and no debris ($N=120$).}
\label{fig2}
\end{figure*}

\noindent \textit{Stellar properties}: Comparing the stellar properties 
of the AAPS sample with the volume limited radial-velocity planet hosts 
sample from \cite{Marshall2014} using the Fisher exact test, we find 
$p$-values in the range 0.5 to 1.0 for their effective temperatures, 
$T_{\rm eff}$, ages and metallicities.  The results are plotted in 
Figure~\ref{fig2}.  The stellar composition of the two samples are 
therefore similar, to be expected as they are both comprised of targets 
from radial velocity planet searches.  However, a comparison of the 
distance distributions shows a marked dissimilarity, with a $p$-value of 
0.057 as the AAPS stars, by and large, lie beyond 20~pc.  The dust 
upper-limits are a strong function of the stellar distance with only 
weak constraints on the presence of debris beyond 25~pc (in the range 
$10^{-5}$ to $10^{-4}$), but the $p$-value for dust incidence comparison 
of the AAPS sample presented here with \cite{Eiroa2013} is 0.599, so the 
two samples are indistinguishable regarding the presence of debris.

\noindent \textit{Planetary mass (limits)}: Stars with sub-Jovian mass 
limits are preferentially low (sub-Solar) metallicity, suggesting that 
these stars do not have high mass planetary companions, as seen in 
\cite{Marshall2014}.  The significance of this correlation is low, as 
the low metallicity stars in the sample are older and closer than the 
average for the sample, such that bias plays a role there.

\noindent \textit{Debris brightness}: Looking at the distribution of 
detected debris discs in the sample, cool Jupiter host stars seem to 
favour the presence of a bright debris disc over low mass planet host 
stars, a trend identified by \cite{Maldonado2012}.  Checking this 
correlation in our sample with the KS test, we obtain a $p$-value of 
0.17 which is suggestive, but not conclusive, of a correlation between 
cool Jupiter mass planets and debris.  The strength (weakness) of the 
correlation is heavily influenced by the presence of HD~207129, a large, 
bright debris disc host star \citep{Krist2010,Marshall2011,Loehne2012}, 
in the potential low-mass planet hosts subgroup.  If we omit this star, 
then the $p$-value of the KS test drops to 0.05, strengthening the 
significance of the (potential) correlation.

\subsection{Conclusions}

We have analysed a sample of AAPS stars, combining upper limits (ruling 
out Jovian analogues around several stars) with radial velocity 
detections, and have been able to identify a weak trend of debris 
brightness with planet mass.  Further analysis, searching for other 
previously identified trends, is hampered by the weak upper limits on 
the presence of debris due to the larger distances to most of the stars 
in our sample than those of other samples, which typically concentrate 
on nearby stars ($d < 25 $pc).

The absence of any strong trends between planets and debris may be a 
function of the dynamical history of these systems wherein the chaotic 
dynamical evolution of planets (including migration and scattering) 
dominates the observed disc brightness.  Any correlations visible in 
more strictly defined stellar samples, wherein we have a better 
understanding of the relative incidences of their component parts as a 
function of the stellar and planetary properties is thereby diluted.

%--------------------------------------------------------------------
\acknowledgements

This research is supported by Australian Research Council grants 
DP0774000 and DP130102695.  JPM is supported by a UNSW Vice-Chancellor's 
Fellowship.  We have made use of NASA's Astrophysics Data System (ADS), 
and the SIMBAD database, operated at CDS, Strasbourg, France.  This 
research has also made use of the Exoplanet Orbit Database and the 
Exoplanet Data Explorer at exoplanets.org \citep{wright11}.

%--------------------------------------------------------------------

%----------------------------------------------------------
%  added 76 stars to cover all in Bond08 paper. total=141

\begin{deluxetable}{lrrrrr}
\tabletypesize{\scriptsize}
\tablecolumns{6}
\tablewidth{0pt}
\tablecaption{Summary of Radial-Velocity Data}
\tablehead{
\colhead{Star} & \colhead{$N$} & \colhead{$\Delta T$ (days)} & 
\colhead{RMS (\ms)} & \colhead{Telescope} & \colhead{Reference} }
\startdata
\label{rvdata}
GJ 729 & 30 & 3218 & 20.7 & AAT & \\
GJ 832 & 39 & 5465 & 5.7 &  AAT & \citet{gj832} \\
       & 54 & 1089 & 1.9  &  HARPS & \citet{gj832} \\
       & 16 & 818 & 1.7 & PFS & \citet{gj832} \\
total  & 109 & 5569 & 3.5 &    & \\ 
HD 142 & 86 & 5667 & 11.3 & AAT &  \\
HD 1581 & 110 & 5668 & 3.3 &  AAT & \\
HD 2039\tablenotemark{a} & 46 & 4780 & 14.0 & AAT & \\
HD 3823\tablenotemark{a} & 75 & 5668 & 5.6 & AAT & \\
HD 4308 & 41  & 680 & 1.7 &  HARPS & \citet{udry06} \\
        & 109 & 5669 & 3.8 & AAT & \\ 
total   & 150 & 5669 & 3.3 &  & \\
HD 7570 & 53 & 5665 & 6.3 &  AAT &  \\
HD 9280\tablenotemark{a} & 30 & 4970 & 12.6 & AAT & \\
HD 10360 & 64 & 5668 & 4.7 &  AAT &  \\
HD 10647 & 51 & 5134 & 10.7 &  AAT &  \\
HD 10700 & 248 & 5726 & 3.0 &  AAT & \\ 
         & 638 & 8800 & 7.1 & Lick & \citet{fischer14} \\
total    & 886 & 9511 & 6.2 &   & \\
HD 11112\tablenotemark{a} & 37 & 5724 & 15.8 & AAT & \\
HD 12387\tablenotemark{a} & 25 & 4403 & 8.0 & AAT & \\
HD 13445 & 64 & 5724 & 5.9 &  AAT & \\
HD 14412 & 26 & 2561 & 3.4 & AAT & \\
HD 16417 & 117 & 5724 & 3.8 &  AAT & \\
HD 17051 & 36 & 4843 & 18.0 & AAT & \\
HD 18709\tablenotemark{a} & 23 & 5104 & 8.5 &  AAT & \\
HD 19632\tablenotemark{a} & 30 & 3863 & 24.8 &  AAT & \\
HD 20201 & 31 & 5105 & 8.0 &  AAT & \\
HD 20766 & 50 & 5881 & 6.5 &  AAT & \\
HD 20782 & 53 & 5520 & 5.8 & AAT & \\
HD 20807 & 91 & 5724 & 4.4 &  AAT & \\
HD 23079 & 37 & 5132 & 5.6  &  AAT & \\
HD 23127 & 44 & 4850 & 11.6 & AAT & \\
HD 23484 & 19 & 2976 & 14.0 & AAT & \\
HD 26965 & 104 & 3046 & 4.4 &  AAT & \\
         & 78 & 5016 & 7.8 & Lick & \citet{fischer14} \\
total    & 182 & 6941 & 6.1 &  &  \\
HD 27274 & 28 & 4114 & 7.0 &  AAT & \\
HD 27442 & 96 & 5724 & 7.3 & AAT & \\
HD 30177 & 36 & 5438 & 18.5 & AAT & \\
HD 30295\tablenotemark{a} & 33 & 4850 & 9.2 & AAT & \\
HD 31827\tablenotemark{a} & 29 & 5265 & 8.1 &  AAT & \\
HD 33811\tablenotemark{a} & 26 & 4878 & 8.8 &  AAT & \\
HD 36108\tablenotemark{a} & 34 & 5549 & 4.0  & AAT & \\
HD 38283\tablenotemark{a} & 64 & 5883 & 4.0 &  AAT & \\
HD 38382\tablenotemark{a} & 45 & 5936 & 5.4 &  AAT & \\
HD 38973\tablenotemark{a} & 43 & 5882 & 5.2 &  AAT & \\
HD 39091 & 69 & 5879 & 6.4 & AAT & \\
HD 40307 & 28 & 5882 & 5.8 &  AAT & \\
         & 345 & 1912 & 1.1 &  HARPS & \citet{tuomi13} \\
total    & 373 & 5882 & 1.9 & & \\ 
HD 42902\tablenotemark{a} & 17 & 4840 & 24.6 &  AAT & \\
HD 43834 & 131 & 5880 & 6.1 &  AAT & \\
HD 44120\tablenotemark{a} & 39 & 5882 & 3.8 &  AAT & \\
HD 44594\tablenotemark{a} & 43 & 5937 & 6.8  & AAT & \\
HD 45289\tablenotemark{a} & 34 & 5941 & 5.1  & AAT & \\
HD 52447\tablenotemark{a} & 24 & 3689 & 15.8 & AAT & \\
HD 53705\tablenotemark{a} & 130 & 5880 & 4.3  & AAT & \\
HD 53706\tablenotemark{a} & 45 & 5881 & 3.3 & AAT & \\
HD 55720\tablenotemark{a} & 28 & 5634 & 3.8 & AAT & \\
HD 59468\tablenotemark{a} & 45 & 5881 & 5.0 & AAT & \\
HD 69655 & 30 & 4754 & 5.6 & AAT & \\
HD 69830 & 19 & 1181 & 4.4 &  AAT & \\
         & 32 & 3451 & 7.8 & Lick & \citet{fischer14} \\
total    & 51 & 4848 & 6.7  &  \\
HD 70642 & 41 & 5882 & 4.4  & AAT & \\
HD 72769\tablenotemark{a} & 30 & 5637 & 4.8 & AAT & \\
HD 73121\tablenotemark{a} & 43 & 5961 & 5.9 & AAT & \\
HD 73524\tablenotemark{a} & 84 & 5935 & 5.4 &  AAT & \\
HD 73526 & 36 & 5226 & 7.7 & AAT & \citet{hkpaper} \\
         & 20 & 856  & 2.8 & PFS & \citet{hkpaper} \\
total    & 56 & 5226 & 6.3  &  & \\
HD 75289 & 46 & 5879 & 6.6 & AAT & \\
HD 76700 & 43 & 4785 & 6.4 &  AAT & \\
HD 78429\tablenotemark{a} & 38 & 5788 & 8.6 & AAT & \\
HD 80635\tablenotemark{a} & 23 & 4784 & 10.6 &  AAT & \\
HD 83443\tablenotemark{a} & 23 & 1211 & 10.2 & AAT & \\
HD 83529A\tablenotemark{a} & 31 & 5964 & 4.9  & AAT & \\
HD 86819\tablenotemark{a} & 34 & 5844 & 10.2 & AAT & \\
HD 88742\tablenotemark{a} & 35 & 5941 & 12.3 & AAT & \\
HD 92987\tablenotemark{a} & 51 & 5935 & 5.3 & AAT & \\
HD 93385\tablenotemark{a} & 45 & 5845 & 7.8 & AAT & \\
HD 96423\tablenotemark{a} & 38 & 5464 & 5.1 & AAT & \\
HD 100623 & 95 & 3305 & 3.7 & AAT & \\
HD 102117 & 59 & 5766 & 4.7 & AAT & \\
HD 102365 & 178 & 5881 & 2.7 & AAT & \\
HD 102438 & 53 & 5881 & 4.1 & AAT & \\
HD 103932 & 18 & 2978 & 5.5 & AAT & \\
HD 105328 & 52 & 5961 & 6.2 & AAT & \\
HD 106453 & 36 & 3014 & 10.7  & AAT & \\
HD 108147 & 56 & 5166 & 13.0 & AAT & \\
          & 118 & 1076 & 16.0 & CORALIE & \citet{pepe02} \\
total     & 174 & 5166 & 15.0 &  & \\
HD 108309 & 66 & 5961 & 4.7 & AAT & \\
HD 114613 & 235 & 5965 & 3.9  & AAT & \\
HD 114853 & 57 & 5962 & 6.9 & AAT & \\
HD 115617 & 153 & 3228 & 3.1  & AAT & \\
          & 78 & 1682 & 2.3 & Keck & \citet{vogt10} \\
total     & 231 & 3228 & 3.1 &  & \\ 
HD 117618 & 73 & 5881 & 5.7 & AAT & \\
HD 120690 & 11 & 1176 & 4.5 & AAT & \\
HD 122862 & 100 & 5961 & 4.7 & AAT & \\
HD 128620 & 102 & 4926 & 3.5  & AAT & \\
HD 128621 & 119 & 5725 & 3.7 & AAT & \\
HD 134060 & 95 & 5876 & 6.3 & AAT & \\
HD 134330 & 44 & 5549 & 6.2 & AAT & \\
HD 134987 & 73 & 5579 & 3.0 & AAT & \\
HD 140901 & 113 & 5551 & 12.1 & AAT & \\
HD 142415 & 22 & 2687 & 17.0  & AAT & \\
          & 137 & 1529 & 14.6 & CORALIE & \citet{mayor04} \\
total     & 159 & 3808 & 14.8  &  &  \\
HD 143114 & 40 & 5878 & 6.2 & AAT & \\
HD 147722 & 66 & 5879 & 17.8 & AAT & \\
HD 154857 & 42 & 4109 & 3.2  & AAT & \citet{2jupiters} \\
HD 155974 & 50 & 5878 & 9.6  & AAT & \\
HD 159868 & 49 & 4077 & 6.6  & AAT & \citet{142paper} \\
          & 34 & 1593 & 4.4  & Keck & \citet{142paper} \\
total     & 83 & 4077 & 5.8 &  & \\
HD 160691 & 172 & 5581 & 2.6  & AAT & \\
          & 40 & 2483 & 7.7 & CORALIE & \citet{pepe07} \\
          & 86 & 980 & 1.7  & HARPS & \citet{pepe07} \\
total     & 298 & 5581 & 3.5  &  & \\
HD 161612 & 50 & 5874 & 4.4 & AAT & \\
HD 164427 & 44 & 5079 & 6.2 & AAT & \\
HD 177565 & 165 & 3501 & 3.1 & AAT & \\
HD 179949 & 68 & 5085 & 12.2  & AAT & \\
HD 183877 & 43 & 5673 & 5.8 & AAT & \\
HD 187085 & 69 & 5434 & 12.0 & AAT & \\
HD 189567 & 88 & 5345 & 5.5 & AAT & \\
HD 191408 & 177 & 5497 & 3.9 & AAT & \\
HD 192310 & 158 & 5377 & 3.1 & AAT & \\
HD 192865 & 44 & 5345 & 10.4 & AAT & \\
HD 193193 & 52 & 5652 & 5.8 & AAT & \\
HD 193307 & 79 & 5377 & 4.4 & AAT & \\
HD 194640 & 78 & 5377 & 4.8 & AAT & \\
HD 196050 & 55 & 5021 & 7.7 & AAT & \\
HD 196068 & 35 & 5375 & 11.8 &  AAT & \\
HD 196761 & 45 & 3010 & 5.7 & AAT & \\
HD 196800 & 38 & 5675 & 6.6 & AAT & \\
HD 199190 & 52 & 5521 & 4.8 & AAT & \\
HD 199288 & 80 & 5521 & 5.1 & AAT & \\
HD 199509 & 32 & 5732 & 4.6 & AAT & \\
HD 202628 & 30 & 4367 & 10.9 & AAT & \\
HD 204385 & 38 & 5465 & 6.5  & AAT & \\
HD 205536 & 27 & 5432 & 4.1 & AAT & \\
HD 205390 & 33 & 5521 & 9.6 & AAT & \\
HD 207129 & 120 & 5433 & 4.9 & AAT & \\
HD 207700 & 33 & 5522 & 5.4 & AAT & \\
HD 208487 & 46 & 5433 & 8.5 & AAT & \\
HD 208998 & 34 & 5054 & 7.8 & AAT & \\
HD 209653 & 40 & 5462 & 5.0  & AAT & \\
HD 210918 & 68 & 5521 & 5.0 & AAT & \\
HD 211317 & 41 & 5433 & 5.1  & AAT & \\
HD 212168 & 47 & 5464 & 5.4 & AAT & \\
HD 212330 & 31 & 5218 & 3.7 & AAT & \\
HD 212708 & 35 & 4069 & 4.3 & AAT & \\
HD 213240 & 35 & 4487 & 5.0  & AAT & \\
HD 214759 & 30 & 5465 & 5.7 & AAT & \\
HD 214953 & 78 & 5464 & 4.2  &  AAT & \\
HD 216435 & 74 & 4723 & 6.5  & AAT & \\
HD 216437 & 51 & 5668 & 4.6  & AAT & \\
          & 21 & 865 & 8.0 & CORALIE & \citet{mayor04} \\
total     & 72 & 5668 & 5.8  &  &  \\
HD 217958 & 35 & 4727 & 9.2 & AAT & \\
HD 219077 & 69 & 5635 & 4.8 & AAT & \\
HD 220507 & 27 & 5464 & 4.2 & AAT & \\
HD 221420 & 79 & 5960 & 4.0 & AAT & \\
HD 222237 & 30 & 5431 & 4.6  & AAT & \\
HD 223171 & 58 & 5464 & 6.0  & AAT & \\
\enddata
\tablenotetext{a}{No far-IR observations available.}
\end{deluxetable}

%----------------------------------------------------------

%----------------------------------------------------------

%----------------------------------------------------------

\end{document}